# RECONSTRUCTION OF THE LARGE SCALE STRUCTURE

Y. HOFFMAN

Racah Institute Of Physics, Hebrew University, Jerusalem, Israel.

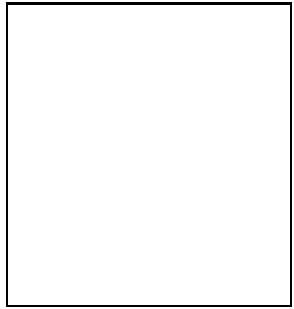

**Abstract**

The large scale structure of the universe is reconstructed by means of the Wiener filter and constrained realizations of Gaussian fields. The density field constructed from optically selected galaxies (Hudson [7]) has been sampled out to a distance of $80h^{-1}Mpc$. This data set is used for the reconstruction of the underlying primordial perturbation field. A preliminary reconstruction analysis from the IRAS $1.9Jy$ density field is presented here for the sake of comparison. The perturbation field within a cube of $160h^{-1}Mpc$ on the side, centered on the Local Group, and with Gaussian smoothing of length $10h^{-1}Mpc$ is reconstructed, assuming the standard CDM model. In particular, the density field at the Zone of Avoidance is studied. A preliminary maximum likelihood analysis performed on Hudson's data shows that it is consistent with a tilted CDM with $n \sim -0.5 \pm 0.3$ and $\sigma_8 \sim 1.4 \pm 0.15$

## 1  Introduction

A problem one often encounters in many branches of physics and technology is the reconstruction of an underlying random field from sparse and incomplete measurements or observations, which are in general subject to statistical errors and are obtained with a finite resolution. One particular case is that of the reconstruction of the large scale distribution of galaxies from galaxy surveys such as the IRAS red-shift survey [18]. The underlying smooth density field is sampled by the discrete galaxy distribution which always suffers from an incomplete sky coverage, e.g. due to the obscuration by the Galactic Plane (the 'Zone of Avoidance') or due to incomplete survey. It is highly desirable to have an efficient algorithm that 'cleans' the statistical (shot noise) errors and extrapolates the field to unsurveyed regions. The outcome of such a method can serve to predict, for example, the large scale structure that lies behind the Zone of Avoidance or to set initial conditions for N-body simulations. Mathematically speaking, the statistical unbiased errors constitute a shot (or white) noise and the blurriness introduced by the finite accuracy of the observations is represented by a point spread function (PSF). In the cosmological case the discreteness of the galaxy distribution introduces Poisson 'shot-noise' and the incomplete sky coverage acts like a convolution of a 'window' function with the underlying field, *i.e.* it acts like a PSP. The act of reconstruction therefore involves a deconvolution.

However, it is well known that in the presence of noise a straight forward deconvolution is highly unstable, and a regularized inversion is to be used.

In most physical problems of deconvolution the number of degrees of freedom of the underlying field is much larger than the number of observed data points, and the underlying field is not completely determined by the data. Thus many different realizations, all of which are consistent with the data, can be constructed. Given that, the problem of reconstruction can be handled in two possible ways. The simplest one is of finding an optimal estimator of the field, and it is usually handled within the framework of maximum entropy method [15], or by the very efficient and more physical approach of the Wiener filter ([14], [16]). However, the estimator corresponds to a mean field, given some theoretical assumptions, and it is much smoother than realizations that reproduce the observable data. This brings us to the other way of generating typical realizations of the underlying field that are consistent with the observations. This can be done only within a framework where the probability distribution function (PDF) is specified *a priori*, hereafter this is referred to as a prior. In the cosmological framework the PDF is assumed to be Gaussian.

Given the PDF the problem turns out to be that of conditional probability. In this framework one can calculate the ensemble mean field, given the observed data, and the statistical properties of the residual from the mean. Moreover, random realizations of the field which are constrained by the data can be reconstructed (Bertschinger [1], Binney and Quinn [3], and Hoffman and Ribak [8]). All of these authors suggested various algorithms for making constrained realizations (CR) of Gaussian fields, where the constraints are assumed to be known exactly without any observational uncertainties. The simple algorithm of Hoffman and Ribak [8] enables the construction of realizations where large number of constraints ($\gtrsim 10^3$) are imposed. The method has been applied recently to the reconstruction of the large scale structure from the velocity potential as produced by the POTENT algorithm [4]. Ganon and Hoffman [6] extended the constrained realizations (CR) formalism to include statistical errors and used POTENT data to make realizations of the density field assuming the standard CDM model [2]. We start here with a brief review of the Wiener filter estimator and its relation to CR of Gaussian fields (§2). The algorithm of CR is applied in §3 to the density field of optically selected galaxies (Hudson [7]) and the density field of the IRAS galaxies (Strauss *et al.* [18]). Here we focus on reconstructing the structure in the Zone of Avoidance, *i.e.* structure obscured by the Galactic disk (§4). A preliminary likelihood analysis is performed on the data set of the optical galaxies, to find the power spectrum that is most compatible with the data (§6). A general discussion is given in §7.

## 2 Constrained realizations and the Wiener filter

Suppose that an underlying random field has $N$ degrees of freedom, and it is specified by $a_\mu$ ($\mu = 1, ..N$) in some functional representation. The observational data set $\{c_i\}$ ($i = 1, ..M$) is related to the field by

$$c_i = W_{i\mu} a_\mu + \epsilon_i, \qquad (1)$$

where $W_{i\mu}$ is the operator which represents the PSF, and $\epsilon_i$ is the statistical error associated with the observed $c_i$ and it is assumed to be drawn from a 'white noise'. We start first with the more restrictive case in which the random field $a_\mu$ is assumed to be Gaussian. The PDF of such a field is determined by its correlation matrix (*i.e.* power spectrum) $\langle a_\mu a_\nu \rangle$, where $\langle ... \rangle$ denotes an ensemble average. The conditional PDF of the field given the data is:

$$P(\vec{a}|\vec{c}) = \frac{P(\vec{a}, \vec{c})}{P(\vec{c})} \qquad (2)$$

In the case of Gaussian fields the ensemble conditional mean field equals to the most probable field given the data, and it is given by:

$$a_\mu^{\rm mp} = \langle a_\mu c_i \rangle \{\langle c_i c_j \rangle\}^{-1} c_j \qquad (3)$$

This yield,
$$a_\mu^{\mathrm{mp}} = W_{i\alpha}\langle a_\mu a_\alpha\rangle \{W_{i\alpha'}W_{j\nu}\langle a_{\alpha'}a_\nu\rangle + \sigma^2\delta_{ij}\}^{-1}c_j, \qquad (4)$$
and here for the sake of simplicity we assumed uncorrelated errors $\langle \epsilon_i\epsilon_j\rangle = \sigma^2\delta_{ij}$.

Once the mean field is computed a CR is easily constructed at a negligible computational cost by adding the residual to the mean field as described by Hoffman and Ribak [8]:
$$a_\mu = \tilde{a}_\mu + \langle a_\mu c_i\rangle\{\langle c_i c_j\rangle\}^{-1}(c_j - W_{j\nu}\tilde{a}_\nu) \qquad (5)$$
(here $\tilde{a}_\mu$ is a random realization of the field.) A somewhat different method for calculating the most probable field within the framework of Gaussian fields is given by Stebbins [17]. It provides an interesting Baysian approach but it is much less efficient than the expression given by Eq. 4.

In the case where the PDF of the underlying field is unknown the variance of the residual from the mean is undetermined. However, assuming the covariance function (matrix) of the field is known an optimal estimator of the field, given the data, can be constructed by the method of the Wiener filter (see [14] and the excellent review of the subject by Rybicki and Press [16]). Given the relation of Eq. 2, one asks for an estimator of the underlying field which is optimal in the sense of minimal variance from the actual field. A linear dependence on the observable $\vec{c}$ is assumed,
$$\vec{a}^{\mathrm{est}} = \bar{\bar{F}}\vec{c} \qquad (7)$$
where matrix notation is used here. One solves for $\bar{\bar{F}}$ by minimizing the discrepancy of $\vec{a}^{\mathrm{est}}$ from the actual underlying field $\vec{a}$. This was solved first for the simple case of an orthogonal representation where the correlation function is diagonal [14] and more recently for a general covariance function but without s PSF [16]. This can be easily extended to the case of an arbitrary PSF. Thus one looks for an $\bar{\bar{F}}$ which minimizes $\langle(\vec{a} - \vec{a}^{\mathrm{est}})^2\rangle$. Extending Eqs. 4-7 of Rybicki and Press [16] yields:
$$\bar{\bar{F}} = \langle \vec{a}\otimes\vec{c}\rangle\langle\vec{c}\otimes\vec{c}\rangle^{-1} = \langle\vec{a}\otimes\vec{a}\rangle\bar{\bar{W}}^\dagger\{\bar{\bar{W}}\langle\vec{a}\otimes\vec{a}\rangle\bar{\bar{W}}^\dagger + \sigma^2\bar{\bar{I}}\}^{-1}. \qquad (8)$$

The important result found here is that the optimal estimator of the field which is obtained by the Wiener filter equals to the most probable (and the ensemble average given the data) field in the case of Gaussian fields. Note that the estimator of the filter minimizes the $\chi^2$ with respect to the underlying field, where $\chi^2 = c_i\{\langle c_i c_j\rangle\}^{-1}c_j$.

To summarize, the Wiener filter yields the optimal estimator of the underlying field, in the sense of minimal $\chi^2$. Taking the step of assuming the random field to be Gaussian, the minimal $\chi^2$ solution corresponds to a maximum likelihood solution and the statistics of the residual from the mean is determined. Furthermore, it can be shown that the minimal variance solution equals to the one obtained by the maximum entropy method (Stebbins [17]). The CR method provides realizations of that residual that are constrained by the data. The main virtue of the CR is in setting typical field configurations (or any linear functional) of the underlying field for Monte Carlo (say) simulations. In cosmology these can be used to set initial conditions for N-body simulations. Elsewhere, we have applied the Wiener filter to recover the angular galaxy distribution from the spherical harmonics decomposition of the IRAS 1.2$JY$ catalog ([11], see also Lahav [10]).

## 3 Large scale structure: data

The large scale structure of the universe has been thoroughly investigated in recent years by mapping the galaxy distribution and independently the velocity field. Within the framework of gravitational instability the comparison of the two fields yields valuable information and constraints on the mass distribution (Dekel *et al.* [5]). The canonical model of cosmology assumes that the large scale structure of the universe has emerged out of a primordial density perturbation field. The amplitude of the field is assumed to be small on relevant scales at earlier times, so that its dynamics follows the linear gravitational instability. It is further assumed that in the linear regime it is a Gaussian random field.

At present, the structure constitutes a non-linear deviation from a uniformly expanding universe, however on scales larger than $\sim 10h^{-1}Mpc$ the deviations are still linear. Thus on scales large enough one can use the Wiener filter approach to find optimal estimator of the primordial perturbation field and CR can be made to make typical realizations of that field which are fully compatible with existing data. These are based on making the *a priori* assumption on the statistical nature of the primordial field, which for Gaussian field means determining its power spectrum. Now, one of the major goals of cosmology is the determination of the power spectrum and currently no model seems to be consistent with all observational constraints. Here, for the lack of any better one the so-called standard cold dark matter (CDM) model is assumed with $\Omega_0 = 1.0$ and $h = 0.5$ (where $\Omega_0$ is the density parameter and $h$ is Hubble's constant in units of $100 Km/s/Mpc$).

Before proceeding to describing the work presented here, a cautionary remark has to be made on the application of the linear theory. Strictly speaking the linear theory is incompatible with the present large scale structure of the universe because of non-linear fluctuations on small scales. A common practice however is to smooth the field so as to remove non-linear short waves, and often a Gaussian filter of a smoothing length of $10 - 12h^{-1}Mpc$ is used ([7], [18], [4]). However the resulting smoothed field does not fully recover the linear field and its Gaussian nature, although it is quite close to it. In most of the work described below the smoothed field is assumed to be linear and Gaussian. A better approach is to use a 'time machine' to take the field back in time (Nusser and Dekel [13]). Preliminary application of this machine shows that for more quantitative studies, such as a likelihood analysis, this correction is quite essential.

Our main purpose here is to use the smoothed density field of optically selected galaxies of Hudson ([7], hereafter referred to as ZCAT), as a data base for CR. This is part of a wider project where we analyze the statistical properties of the ZCAT and IRAS density fields, the peculiar velocity field and the relations between these fields. In particular the CR are used to 'predict' the structure that is hidden behind the Zone of Avoidance. For the sake of comparison with ZCAT, we have also used the IRAS $1.9Jy$ density field ([18]) as a data base. The realizations constitute also an estimator of the underlying field that has been cleaned from the shot noise.

A detailed description of ZCAT is given by Hudson ([7]), and therefore only aspects relevant to the CR are reviewed here. The ZCAT compilation is based on the UGC and ESO optical catalogs, which cover only 67% of the sky, mostly due to the obscuration by the Galactic disk. The ZCAT density field is evaluated with a Gaussian filter with a smoothing length of $R_s = 10h^{-1}Mpc$. It is constructed first in red-shift space and then transformed to real space by the iterative scheme of Yahil *et al.* [19]. The sparseness of the galaxy distribution introduces shot noise errors whose covariance matrix is evaluated by bootstraping. The ZCAT density field extends out to $80h^{-1}Mpc$ from us. Realizations of the underlying field are made on a cubical $64^3$ grid with a grid size of $5h^{-1}Mpc$. The ZCAT density field is sampled on that grid with a spacing of the smoothing length of $10h^{-1}Mpc$. It is sampled only where ZCAT field is determined from actual data and the unobserved regions are avoided. This produces a data set of some $M \sim 1400$ points, which are used to set constraints on the primordial perturbation field. Note that the sampling spacing equals to the smoothing length and this introduces off-diagonal terms in the error covariance matrix. For the sake of comparison we use the density field constructed from the IRAS $1.9Jy$ catalog as a data base for the realizations. This has been done by taking the IRAS data only where ZCAT data is available, *i.e.* ZCAT mask is imposed on IRAS. The shot noise errors are based on A. Yahil's compilation (private communication). Off-diagonal terms of the errors covariance matrix are neglected in the IRAS case. Note that in general the amplitude of the IRAS' density field is lower than that of ZCAT and therefore it suffers less from non-linear effects.

The CR were calculated by the CONFIGUR program that has been optimized to take full advantage of the vectorizing ability of the CONVEX C-220 computer of the Hebrew University. Currently a realization of the density field on a $64^3$ grid and subject to $\sim 1400$ constraints is done in 11 minutes CPU time. A subroutine performance analysis that was performed on the programs indicates that the calculation of the mean field takes 23% of the CPU time, FFT takes 18% of the time and the matrix inversion takes only 2% of the time.

## 4 Constrained realizations

Standard CDM realizations constrained by ZCAT and IRAS ($1.9 Jy$) density field are presented here. Fig. 1 shows plots of the overdensity and velocity fields superimposed. The upper two plots correspond to the supergalactic plane $SGZ = 0$, and the lower ones show the $SGY = 0$ plane which coincide with the Galactic plane. Thus, the $SGY = 0$ plots yield the reconstructed perturbation field at the Zone of Avoidance. The two left plots are constrained by ZCAT, and the two right ones are based on IRAS sampled at the same points as ZCAT. The overall structure produced by ZCAT and IRAS is generally similar, however there are some marked differences. First, the amplitude of the ZCAT field is higher than IRAS' by roughly a factor of 2. The two reconstructions recover the Perseus-Pisces (PP) supercluster and find a similar structure, yet they differ with respect to the Cetaurus-Hydra complex. The ZCAT map places the peak of that structure at a distance of $\sim 35 h^{-1} Mpc$, compared with the $\sim 45 h^{-1} Mpc$ found from IRAS and which coincides with the Great Attractor (GA) solution [12] (see ref. [7] for detailed discussion).

A more detailed graphical analysis of the structure at the Zone of Avoidance is presented by Figs. 2 and 3 where the overdensity field is plotted in Galactic spherical projection at distances of $R = 40 h^{-1} Mpc$ (Fig. 2) and $R = 60 h^{-1} Mpc$ (Fig. 3). Here we focus on the $b \sim 0°$ region. The $R = 40 h^{-1} Mpc$ IRAS and ZCAT maps are similar and are reproducing the PP and Hydra-Centaurus (or GA) complexes. New features are found at $R = 60 h^{-1} Mpc$. In the ZCAT map the GA almost does not extend beyond this radius and only one extension is found which peaks at $l = 15°$ and $b = -5°$ at that distance. The structure there is statistically robust and it is found in all realization we have made. The proximity of that direction to the Galactic center makes the optical verification of that structure to be very difficult. The IRAS $R = 60 h^{-1} Mpc$ reconstructed map shows a much richer structure, as the GA extends beyond that distance at $l \sim 325°, b = 0°$. A new peak is found here at $l = 285°, b = 5°$, and inspection of Fig. 1 shows it comes from an extension of the GA complex. Interesting, a recent optical redshift survey of that region has found an indication to an excess of galaxies which coincide with the peak reported here (Kraan-Korteweg and Woudt [9]). Again this peak is statistically robust and has been reproduced in all realizations made by us.

## 5 Likelihood analysis

One of the virtues of the CR method is that it can produce a realization regardless of the numerical value of the constraints. A realization that reproduces a $1\sigma$ or $10\sigma$ fluctuation is performed at the same computational efforts. Yet, in applying the method to large data sets so as to reconstruct the large scale structure a major concern is to control the compatibility of the assumed *prior* with observations. This can be easily achieved by performing a likelihood analysis on the data. Note that the calculation of the mean field includes the inversion of the correlation matrix, and therefore the evaluation of the $\chi^2$ can be done with a minimal computational cost. Given that the field is assumed to be Gaussian the likelihood of the data given the theoretical model and the assumed errors, is:

$$L(\text{data}|\text{model}) = \frac{1}{\sqrt{(2\pi)^M \det \langle c_i c_j \rangle}} \exp(-\frac{\chi^2}{2}) \qquad (9)$$

To optimize the power spectrum estimation the approach adopted here is to parametrize the power spectrum and find the optimal parameters by maximizing the likelihood function. The CDM family of models provides a natural choice of parameters, namely $h, \Omega_0$ and the normalization factor which is related to the biasing factor.

A preliminary likelihood analysis is presented here as applied to the 'tilted' CDM family of models models, where the normalization and the power law exponent ($n$) are the free parameter. The normalization constant is related to $\sigma_8^2$ which is the variance within a top-hat sphere of radius $8 h^{-1} Mpc$. The data used here is taken from the ZCAT density field. Special care should be given to the 'linearization' of the observed density field, *i.e.* taking it backwards in time to the linear regime. However, this depends on the assumed biasing factor that relates the ZCAT and mass density fields and on $n$. As a first step in this direction we have applied the Nusser and Dekel [13] 'time machine' to the data assuming a biasing factor of unity and $n = 1$. The algorithm had been applied to the ZCAT data presented on a $32^3$ $5h^{-1}Mpc$ spacing grid, where the unobserved regions have been supplemented with mock data [7]. The periodicity of the FFT limits us to using the linearized data only within an inner $16^3$ cube. In future calculations the procedure will be applied to a $64^3$ grid, thus enabling using the full data set and thereby improving the constraints on the slope of the power spectrum. The linearized field has been then used in the likelihood analysis, ignoring the dependence of the 'linearization' on the normalization. Another problem is how the shot-noise errors are affected by this correction. As a first order approximation we scale the errors with the 'linearized' densities, however this deserves a more careful study. Given all the above *caveats* the preliminary results are summarized by $\sigma_8 = 1.4 \pm 0.15$ and $n = -0.5 \pm 0.3$.

## 6  Discussion

The reconstruction of the primordial perturbation field from sparse and noisy data has been made by the method of CR which is related to the Wiener filter. The ZCAT and IRAS density fields were sampled and these data sets have been used to set constraints on realizations of the standard CDM model. The CR and Wiener filter method are found to be very useful tools of reconstruction and currently it can be easily done using $\lesssim 2000$ constraints on $64^3$ grid. The method is used here to clean shot noise errors and reconstruct the structure lying behind the Zone of Avoidance. Indeed, using IRAS data such a structure has been detected at distance of $60h^{-1}Mpc$, $b \sim 5°$ and $l = 285°$. This has been confirmed by an independent discovery of an overdensity of optically selected galaxies at the same region [9].

**Acknowledgements.** I am grateful to V. Bistolas, A. Dekel, M. Hudson, T. Kolatt, O. Lahav, A. Nusser, A. Yahil and S. Zaroubi for their contribution to the present work and many stimulating discussions.

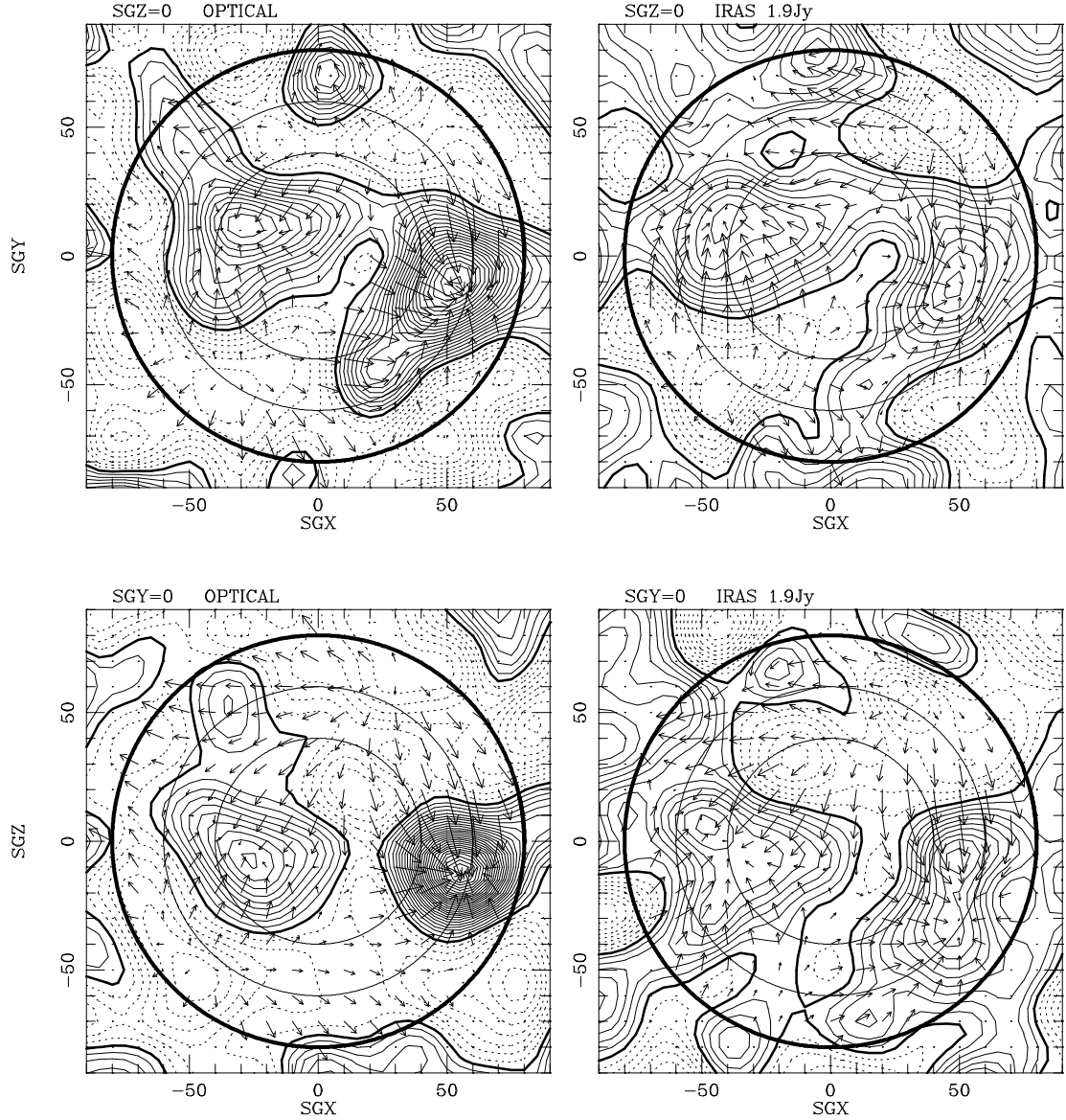

Fig.1: Reconstruction of the density and velocity fields, presented in supergalactic coordinates. The maps correspond to realizations constrained by the ZCAT (leftmaps) and IRAS $1.9Jy$ (right maps) density fields. The upper maps correpsond to the supergalactic ($SGZ = 0$) plane and the lower ones to the Galactic plane which coincides with the $SGY = 0$ plane. The latter ones show the reconstruction of the perturbation field at the Zone of Avoidance. The contour lines represent the fractional overdensity ($\delta$), where the line spacing is 0.1, and the arrows represent the peculiar velocities. Distance unit is $5h^{-1}Mpc$ and the arrows length unit is Hubble's constant times the distant unit.

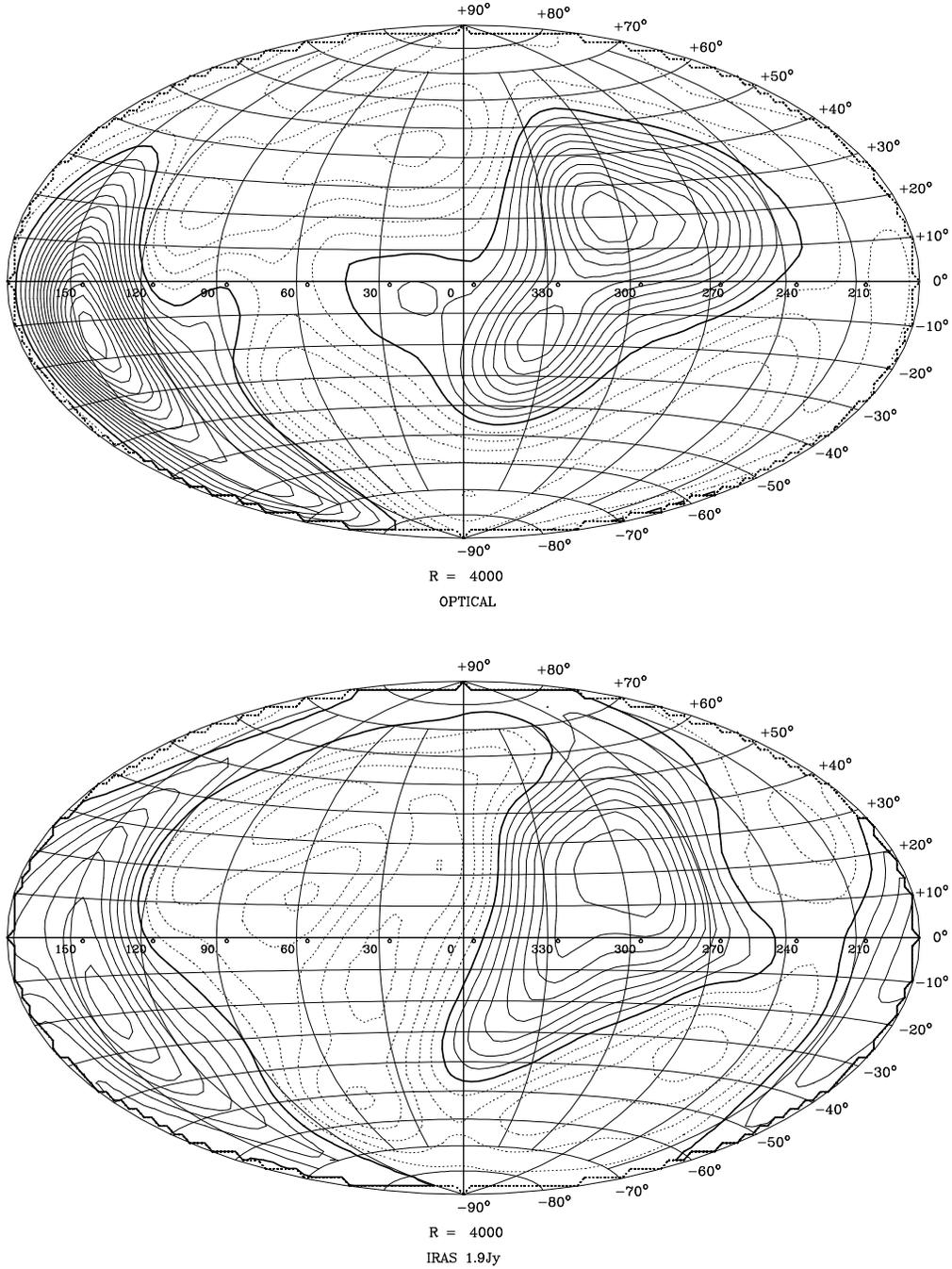

Fig. 2: The reconstructed density field is evaluated on a sphere of radius $40h^{-1}Mpc$ and presented in Galactic coordinates. Upper plot corresponds to the optical data constraints and the lower one to IRAS $1.9Jy$ constraints. Overdensity contours level is 0.1. Note that the galactic center ($l = 0°$) has been shifted to the center of the plots. Both maps are dominated by the GA complex, centered on $l \sim 300°, b \sim 20°$, and the PP supercluster which peaks at $l \sim 150°, b \sim -10°$.

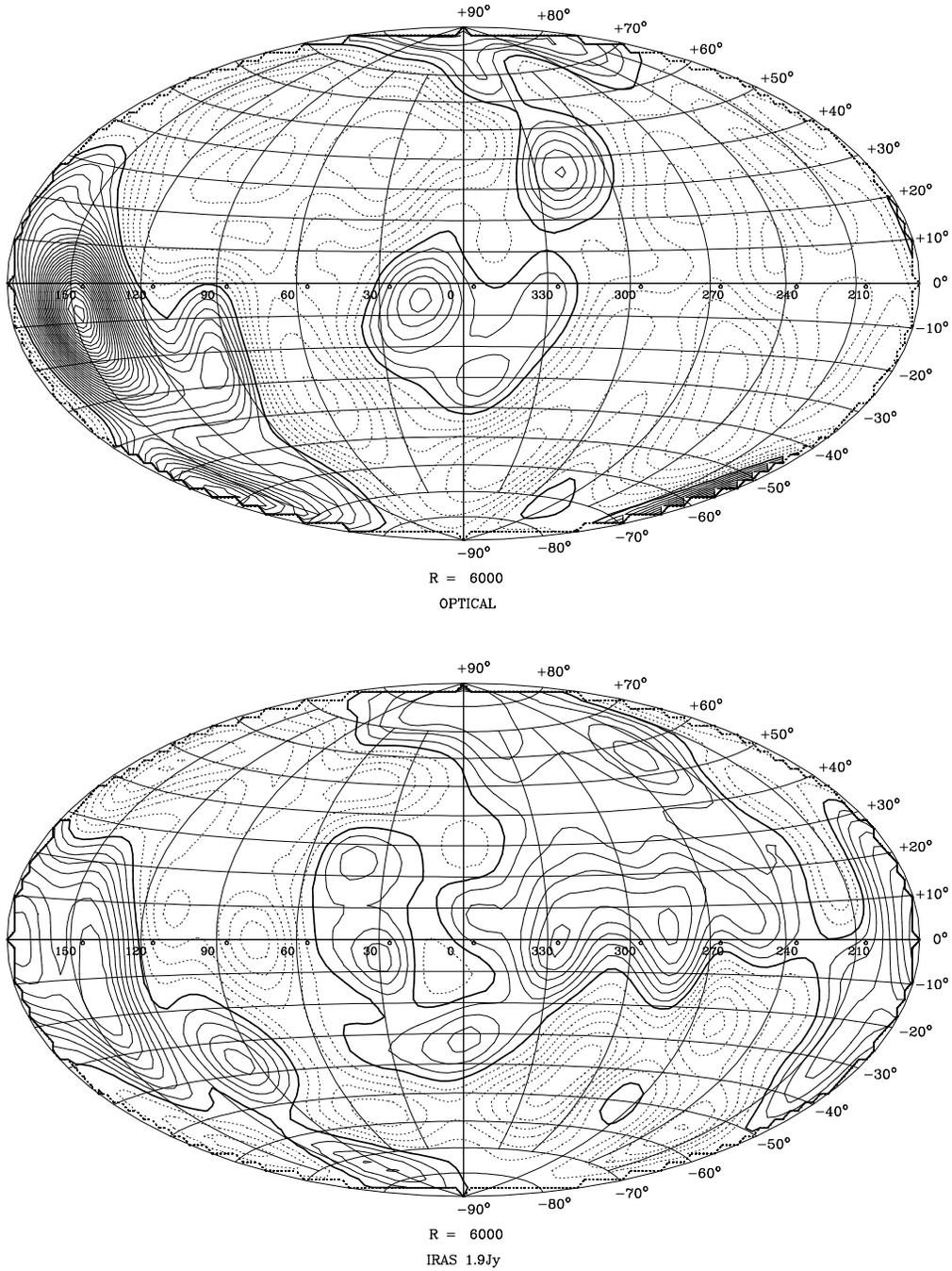

Fig. 3: Same as Fig. 2, but the field is evaluated at a larger radius of $60h^{-1}Mpc$. There is a prediction for a structure in a direction close to the Galactic center in both maps. The IRAS map peak at $l \sim 285°, b \sim 5°$ coincides with a recently discovered excess of galaxies [9].